\documentclass[useAMS, usenatbib,letterpaper,10pt]{mn2e}
\title[The stellar and dark matter haloes of Illustris galaxies]{Halo Mass and Assembly History Exposed in the Faint Outskirts:\\ the Stellar and Dark Matter Haloes of Illustris Galaxies}

\author[Pillepich et al.] {Annalisa Pillepich$^1$\thanks{E-mail: apillepich@cfa.harvard.edu},
Mark Vogelsberger$^2$, Alis Deason$^3$\thanks{Hubble Fellow}, Vicente Rodriguez-Gomez$^1$,
\newauthor
Shy Genel$^1$, Dylan Nelson$^1$, Paul Torrey$^1$, Laura V. Sales$^1$,  Federico Marinacci$^4$$^,$$^5$, 
\newauthor
Volker Springel$^4$$^,$$^5$, Debora Sijacki$^6$, and Lars Hernquist$^1$\vspace{2mm}
\\
$^1${Harvard--Smithsonian Center for Astrophysics, 60 Garden Street, Cambridge, MA 02138}\\
$^2${Department of Physics, Kavli Institute for Astrophysics and Space Research, Massachusetts Institute of Technology, Cambridge, MA 02139, USA}\\
$^3${Department of Astronomy \& Astrophysics, University of California Santa Cruz, 1156 High St., Santa Cruz, CA 95064} \\
$^4${Heidelberg Institute for Theoretical Studies, Schloss-Wolfsbrunnenweg 35, D-69118 Heidelberg, Germany}\\
$^5${Zentrum fuer Astronomie der Universitaet Heidelberg, ARI, Moenchhofstr. 12-14, D-69120 Heidelberg, Germany}\\
$^6${Institute of Astronomy and Kavli Institute for Cosmology, University of Cambridge, Madingley Road, Cambridge CB3 0HA, UK}\\
}

\usepackage[bookmarks,bookmarksnumbered,colorlinks=true, citecolor=blue, linkcolor=black]{hyperref}
\usepackage{amsfonts}
\usepackage{amsmath}
\usepackage{amssymb}
\usepackage{color}
\usepackage[american]{babel}
\usepackage{verbatim}
\usepackage{times}
\usepackage{graphicx}
\usepackage{comment}

\def \MSUN{\rm M_{\odot}}
\def \RVIR{R_{\rm vir}}

\def \RHALF{r_{1/2}}

\def \MVIR{M_{\rm vir}}
\def \MHALO{M_{\rm Halo}}
\def \MSH{M_{\rm Stellar ~Halo}}
\def \MS{M_{\rm STARS}}
\def \AS{\alpha_{\rm STARS}}
\def \ADM{\alpha_{\rm DM}}

\def \etal{{\it et al.\ }}

\begin{document}

\maketitle

\begin{abstract}

We use the Illustris Simulations to gain insight into the build-up of the outer, low-surface brightness regions which surround galaxies. 
We characterise the stellar haloes by means of the logarithmic slope of the spherically-averaged stellar density profiles, $\AS$ at $z=0$, and we relate these slopes to the properties of the underlying Dark-Matter (DM) haloes, their central galaxies, and their assembly histories.
We analyze a sample of $\sim$5,000 galaxies resolved with more than $5\times10^4$ particles each, and spanning a variety of morphologies and halo masses ($3\times10^{11} \le\MVIR\lesssim10^{14}\MSUN$).
We find a strong trend between stellar halo slope and total halo mass, where more massive objects have shallower stellar haloes than the less massive ones ($-5.5\pm0.5<\AS<-3.5\pm0.2$ in the studied mass range).
At fixed halo mass, we show that disk-like, blue, young, and more massive galaxies are surrounded by significantly steeper stellar haloes than elliptical, red, older, and less massive galaxies. 
Overall, the stellar density profiles fall off much more steeply than the underlying DM, and no clear trend holds between stellar slope and DM halo concentration. 
However, DM haloes which formed more recently, or which accreted larger fractions of stellar mass from infalling satellites, exhibit shallower stellar haloes than their older analogs with similar masses, by up to $\Delta\AS\sim0.5-0.7$.
Our findings, combined with the most recent measurements of the strikingly different stellar power-law indexes for M31 and the Milky Way, appear to favour a massive M31, and a Milky Way characterised by a much quieter accretion history over the past 10 Gyrs than its companion.

\end{abstract}



\section{Introduction}

Observations of the Milky Way, M31, and other nearby galaxies demonstrate that the bright, central body of both early and late type galaxies is surrounded by an extended, faint envelope of stars \citep[e.g.][]{MartinezDelgado:2010}. 
This is referred to as stellar halo, or intra-cluster light (ICL) for galaxies at the centres of massive galaxy clusters, and comprises both mixed material as well as more organised features in configuration- and phase-space, including stellar streams, shells, tidal tails, globular clusters, and satellite galaxies \citep[e.g.][]{Belokurov:2006,Tal:2009, Atkinson:2013}.
For the Milky Way and M31, the characterisation of the stellar halo can rely upon samples of individually-resolved stars with photometric and spectroscopic observations, and in some cases proper motions measurements (e.g. from surveys like SDSS/SEGUE, RAVE, APOGEE, Gaia, SPLASH, PAndAS, HSTPROMO).
However, star count studies are limited to relatively small distances \citep[e.g. M81, NGC253, NGC55, CenA by][]{Barker:2009, Bailin:2011,Tanaka:2011, Crnojevic:2013}, and direct detections of the outskirts of more distant galaxies are based on either deep observations of the integrated light in projection \citep[][]{Bakos:2012, Abraham:2014, Dokkum:2014}, or the stacking of a large number of shallower images of similar galaxies, both in the Local Universe and at intermediate redshifts \citep[most recently,][respectively]{DSouza:2014, Tal:2011}.
Also at the highest masses, the stellar envelopes around the brightest cluster galaxies (BCG) have been observed with a variety of techniques, e.g. by \cite{Zibetti:2005, Seigar:2007, Donzelli:2011}.
Such observational efforts are motivated by the possibility that the kinematics, metallicities, ages and spatial distributions of halo and ICL stars might encode information about the entire assembly history of the halo they belong to \citep[e.g.][]{Bell:2008, Schlaufman:2009}.

Stellar haloes are indeed thought to be direct evidence of the hierarchical growth of structure in the Cold Dark Matter paradigm, as numerical simulations have been able to reproduce their broad features from the debris of accreted and disrupted satellite galaxies \citep[e.g. among others][]{Bullock:2005, Abadi:2006, Johnston:2008, Font:2011}.
However, from the theoretical viewpoint, quantitative and reliable predictions on how the spatial structure and global kinematics of the stellar haloes 
correlate with halo mass, halo formation time, properties of the central galaxies, and underlying DM distributions are still missing, and the interpretation of the observational data remains only qualitative. 
Because of computational limitations, numerical studies in this context have come mostly in two flavors:
1) via a combination of N-body-only simulations with semi-analytic models and/or stellar tagging techniques to mimic the stellar components \citep{Cooper:2010, Rashkov:2012, Cooper:2013}; 2) or via gravity+gasdynamics simulations of individual highly-resolved galaxies in cosmological context \citep{Abadi:2006, Zolotov:2009, Tissera:2012, Tissera:2013, Tissera:2014, Puchwein:2010}. 
While the former cannot fully capture the differences in the orbital contents of stars and DM, and cannot reproduce the effects that baryonic physics might imprint into the underlying DM distribution \citep[see][]{Bailin:2014}, the latter cannot assess the relevance of their outcome against the possibility of large halo-to-halo variations because of a lack of statistics. 

In this paper, we bridge the gap between the statistical samples of N-body+tagged haloes and the self-consistent realizations of individual galaxies, by using the Illustris simulations. These are a suite of gravity+hydrodynamics simulations of a (106.5 Mpc)$^3$ volume at kpc 
or better resolution, run with the code {\sc arepo} \citep{Springel:2010} and including key physical processes 
that are believed to be relevant for galaxy formation \citep[see][]{Vogelsberger:2014, Vogelsberger:2014b, Genel:2014}. Illustris thereby provides an ideal framework to undertake the characterisation of the properties of stellar haloes on a galaxy-population basis.
In what follows, we focus on the power-law index of the 3D, spherically-averaged density profiles of stars and DM in the outskirts of the simulated galaxies. We quantify to what extent the stellar halo slope, $\AS$, can be used as a ruler to infer the properties of the underlying DM haloes and their assembly histories, also in relation to the properties of the galaxy residing at the centre of the halo potential wells.
Our study is timely both in light of the advent of deeper surface brightness data (through the Hubble Space Telescope HST and medium-sized, ground-based telescopes), and because the characterisation of the low-surface brightness features in large samples of galaxies, simulated or observed, can aid in the interpretation of the Milky Way's stellar halo and vice versa.

The paper is organised as follows: we introduce the adopted numerical simulations, methods and definitions in Section \ref{SEC_SIMS}; our main result about the relationship between stellar halo slope, halo mass and galaxy properties is presented in Section  \ref{SEC_MHALO}; we compare the stellar and the DM density profiles in Section \ref{SEC_DM}; in Section \ref{SEC_ASSEMBLYHISTORY}, we quantify the correlation between stellar halo slopes and the halo assembly histories, and we compare our findings to observations of the Milky Way and M31. We discuss our results and future directions in Section \ref{SEC_DISCUSSION}, and we summarize in Section \ref{SEC_SUMMARY}.


\begin{figure*}
\begin{center}
\includegraphics[width=16.5cm]{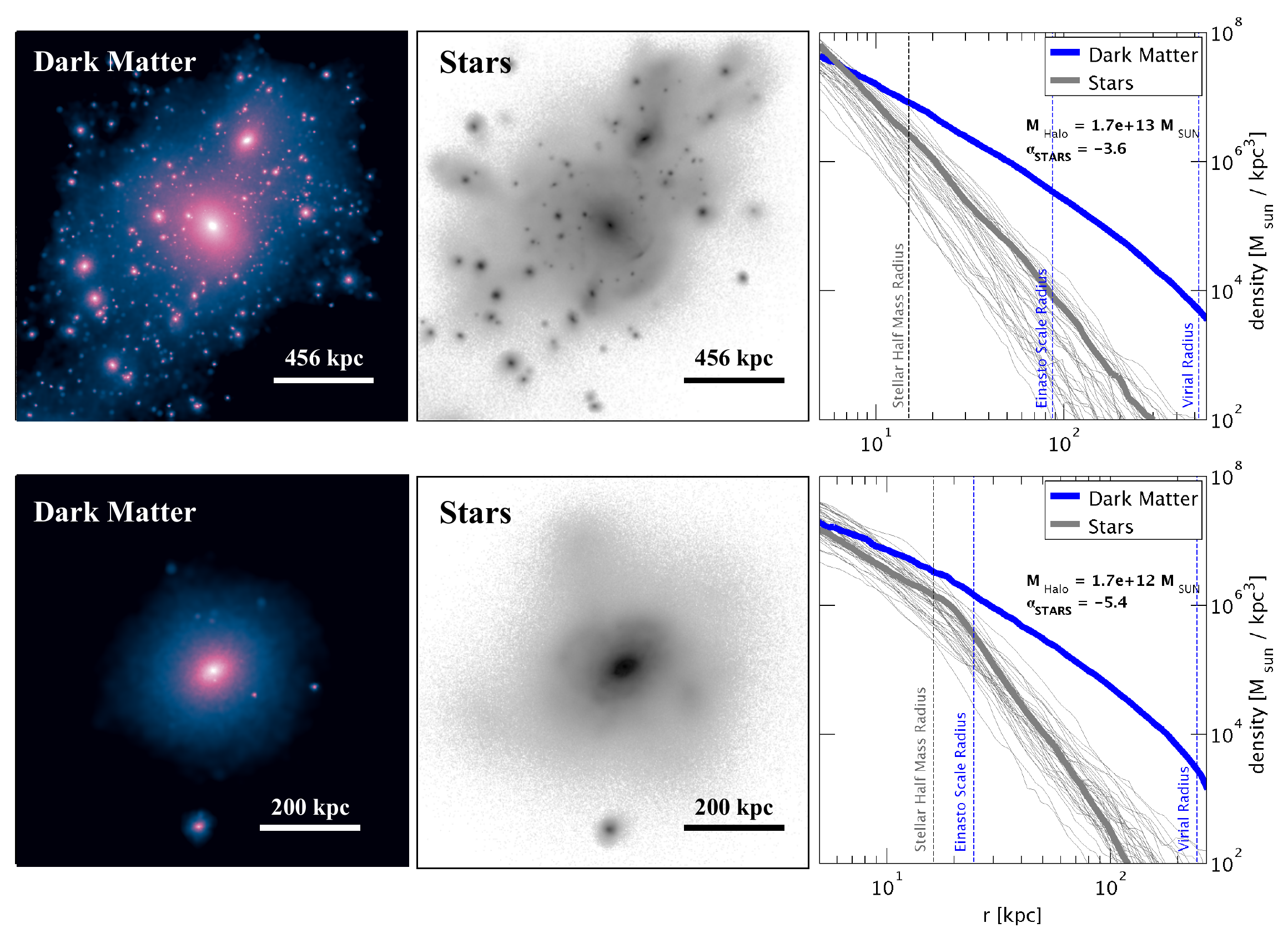}
\caption{2D projections of dark matter density and stellar light (Johnson-K band, see \textcolor{blue}{Torrey \etal 2014} for details) for two galaxies, one elliptical (upper row) and one disk (lower row), with the relevant radial scales and the corresponding 3D spherically-averaged profiles. Even though the density images depict also subhaloes and satellites, these are not accounted for by the fits and profiles on the right-hand side. The thin gray curves in the rightmost panels represent the stellar profiles of a sample of hand-picked Illustris galaxies which have been classified as elliptical (upper panel) and disk galaxies (lower panel) by visual inspection \citep[corresponding to the red and blue dots of Figures \ref{FIG_STELLARSLOPES} and \ref{FIG_GALAXYPROPERTIES}; see][for details]{Vogelsberger:2014b}.}.
\label{FIG_IMAGES}
\end{center}
\end{figure*}

\vspace{-0.5cm}
\section{Simulations and Methods}
\label{SEC_SIMS}

The analysis developed in this paper is based on the Illustris Project (\textcolor{blue}{http://www.illustris-project.org}), a series of gravity+hydrodynamics realizations of a (106.5 Mpc)$^3$ cosmological volume run with the {\sc arepo} code \citep{Springel:2010}. These have been simulated at multiple resolutions, including key physical processes relevant for galaxy formation, and have been recently presented in a series of papers \citep{Vogelsberger:2014, Vogelsberger:2014b, Genel:2014}. 
The highest-resolution run (Illustris-1, or simply Illustris) handles the dark matter component with a mass resolution of $m_{\rm DM} = 6.26\times 10^6 \MSUN$ and the baryonic component with $m_{\rm baryon} \simeq 1.26 \times 10^6 \MSUN$. The co-moving gravitational softening lengths are 1.4 kpc and 0.7 kpc at $z=0$, respectively for DM and baryonic collisionless particles. The gas gravitational softening length is adaptive and set by the cell size, with a floor given by the aforementioned 0.7 kpc, however the sizes of the cells
used to evolve the gas can be much smaller than this.
Two lower resolution simulations of the same volume (Illustris-2 and 3) have mass resolutions 8 and 64 times lower, and softening lengths 2 and 4 times larger. They have all been run from $z=127$ to $z=0$ by adopting cosmological parameters consistent with the latest WMAP-9 results \citep{Hinshaw:2013}.

Illustris has been shown to successfully follow the coevolution of dark and visible matter, by simultaneously reproducing, e.g., the observed evolution of the cosmic star formation rate density and the galaxy stellar mass function from $z = 0$ to $z \sim 6$
\citep{Genel:2014}, the statistics of hydrogen on large scales \citep{Bird:2014}, a reasonable and diverse range of morphologies and colors in the well-resolved galaxy population (\textcolor{blue}{Torrey \etal 2014, to be submitted}), and the observed relationships between galaxies and their central supermassive black holes (\textcolor{blue}{Sijacki \etal 2014, to be submitted}). 
This has been achieved, in part, thanks to the inclusion of: a) primordial and metal-line cooling with self-shielding corrections, b) stellar evolution and feedback processes based on kinetic outflows, c) chemical enrichment modeling nine individual elements, d) a prescription for the metal content of galactic outflows, e) black hole seeding, accretion, and merging processes, f) quasar- and radio-mode feedback, and g) a prescription for radiative electromagnetic feedback from active galactic nuclei \citep[see][for details]{Vogelsberger:2013b, Torrey:2014a}.

Haloes, subhaloes, and their basic properties have been identified with the {\sc fof} and {\sc subfind} algorithms \citep{Davis:1985, Springel:2001, Dolag:2009}, at each of the 136 stored snapshots. 
Herein we use the tree obtained from the newly developed {\sc sublink} code (\textcolor{blue}{Rodriguez-Gomez \etal 2014, in preparation}) to determine halo formation times and the time of a last major merger event (see Section  \ref{SEC_ASSEMBLYHISTORY}).

\subsection{The Galaxy Sample}
\label{SEC_SAMPLE}
Among the identified haloes, we select only galaxies at $z=0$ which are centrals (i.e. not satellites/subhaloes of more massive parents) and resolved by at least 2,000 stellar particles per halo: 
this allows us to identify galactic structure and angular momentum content, and to resolve the star-formation processes within satellite galaxies which are tens of times less massive than their centrals.
In Illustris \citep[whose baryon conversion efficiency at $z=0$ is in broad agreement with observations, see][Figure 12]{Vogelsberger:2014b}, haloes with total mass exceeding $10^{11} \MSUN$ meet these requirements, and make up a sample of 14,000 galaxies, typically resolved by more than $3 \times 10^5$ particles each. 
Nevertheless, our resolution convergence tests (see Section \ref{SEC_MHALO} and Appendix A) show that a more stringent limit is required to properly sample the low-density stellar regions in the outskirts of haloes and thereby measure the stellar profile slope. The results presented in this paper are thus based upon a sample of 4,872 well-resolved galaxies with $\MHALO \geq 3\times 10^{11} \MSUN$ (corresponding to $\MS \sim 3 \times 10^9 \MSUN$)\footnote{Throughout the paper, $\MHALO$ denotes the total mass of DM, gas and stars enclosed within the virial radius ($\RVIR \equiv R_{\rm 200c}$). By stellar mass ($\MS$), we correspondingly mean the sum of all the stellar particles masses contained within the same virial radius, unless otherwise stated. 
}, each resolved with a minimum of 3,300 stars, and more than 54,000 particles in total.

\begin{figure}
\begin{center}
\includegraphics[width=8.5cm]{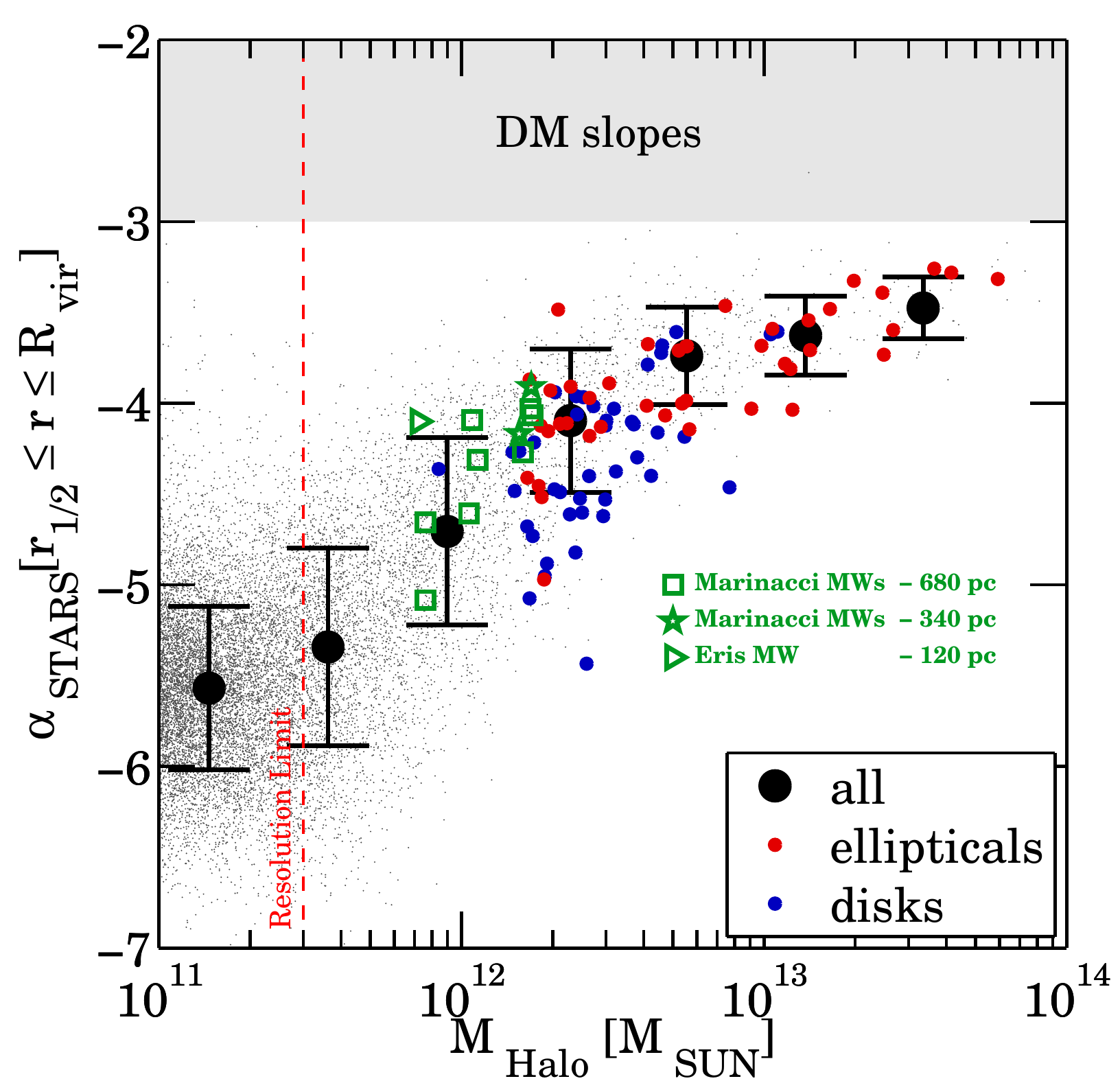}
\caption{The relation between the logarithmic slope of the stellar halo ($\alpha_{\rm STARS}$, calculated between the stellar half-mass radius and the virial radius of the halo) and the total halo mass, at $z=0$. Filled, large black circles denote median values in halo-mass bins for the Illustris galaxies, with corresponding 1-$\sigma$ standard deviations. Green, open symbols represent a sample of individually-simulated, N-body+hydrodynamics zoom-in galaxies \citep[the Aquarius and Eris Milky Way analogs from, respectively, ][]{Marinacci:2014, Rashkov:2013}. Our resolution tests indicate that the flattening of the $\AS-\MHALO$ relation below $3\times 10^{11} \MSUN$ is a numerical artifact, while the amount of scatter is physical. The blue and red dots represent a sample of individual, visually selected galaxies with strong disk (blue) or elliptical (red) morphologies \citep[see][for details]{Vogelsberger:2014b}.}
\label{FIG_STELLARSLOPES}
\end{center}
\end{figure}

\subsection{Fitting Procedures and Definitions}

We characterise the stellar and DM haloes by fitting their 3D spherically-averaged density profiles with a single power-law, $\rho = \rho_0 r^{\alpha}$, within specific radial ranges from the halo centre (defined by the position of the most bound particle).
The profiles are measured for each individual halo, for both DM and stellar particles, in spherical shells evenly spaced in logarithmic radius ($d {\rm log}_{10}(r[{\rm kpc}]) = 0.03$), over the range $\RVIR/50 \leq r \leq \RVIR$. Only particles which the {\sc subfind} algorithm identifies as bound to every given halo are taken into account, thereby after removing the contribution of gravitationally-bound subhaloes and neglecting those particles, occurring especially in the outskirts, that are not linked by the {\sc fof} algorithm to the not-necessarily spherically-symmetric parent halo. 
The fitting procedure is performed in logarithmic space, by minimizing the summed squares of the residuals to a first-degree polynomial fitting function, equally weighting all bins containing at least one particle each. 

For the stellar halo, we are interested in the logarithmic slope or power-law index, $\AS$, of the low-surface brightness stellar component which extends {\it beyond} the main, bright body of a galaxy. Since it is outside the scope of this paper to calculate the photometric decomposition of the surface brightness profile into the different morphological components of a simulated galaxy (e.g bulge, disk, and halo), we simply define the {\it stellar halo} as all the stellar material beyond a given radius. 
We adopt the stellar half-mass radius ($\RHALF$) as the inner boundary of the stellar halo: this varies from $\sim$10 kpc for $10^{12} \MSUN$ objects to 30 kpc for $5\times10^{13} \MSUN$ groups, and thus corresponds to about 3-4 times the disk scale length of Milky-Way like galaxies (see Figure in Appendix A). Although in previous works we have adopted $2 \times \RHALF$ to define a galaxy, observations of Milky Way halo stars can be as close as $\sim$ 10-15 kpc above the disk plane, making the choice quite ambiguous. 
As the outer boundary of the stellar halo, we formally adopt the virial radius; in practice, the slope is measured as long as at least one star falls in the chosen radial bins, which can happen at radii smaller than $\RVIR$ (see Appendix A).

The radial range, over which the halo stellar density is measured, could in principle bias the quoted value of $\AS$. On one hand, the slope can roll and the profile can become steeper in the outer regions of the stellar halo (as it has been suggested by \citealt{Deason:2014a} for the Milky Way, and has been predicted by \citealt{Bullock:2005} via hybrid non-hydrodynamic stellar halo models); on the other hand, the measured slope can be affected by local inhomogeneities and enhancements in phase-space, including stellar shells and streams, making the measurement noisy when calculated across an insufficient radial extent. The key results of the paper are shown for $\AS$ intended as an average rather than a local slope, and measured specifically in the range $\RHALF \leq r \leq \RVIR$; however, we discuss median trends and halo-to-halo variations for a variety of different choices: for example, for comparison with observations, we have also measured the stellar halo slope at fixed apertures, $10 \leq r \leq 50$ kpc and $50 \leq r \leq 100$ kpc, for Milky-Way mass haloes (see Appendix A).

We characterise the shapes of DM haloes in terms of NFW \citep{Navarro:1996, Navarro:1997} and Einasto \citep{Einasto:1965} fits, as well as with methods which are fitting-formula independent and  better able to accommodate the effects that baryons may induce on the underlying DM structures. The fitting is performed via the least squares method, in logarithmic space, and with equally-weighted radial bins. In what follows, we compare the stellar and DM logarithmic slopes, $\AS$ and $\ADM$, in the common range $r_{\rm Einasto} \leq r \leq 1/2 ~\RVIR$, where $r_{\rm Einasto}$ denotes the scale radius in the Einasto profile formula.

In Figure \ref{FIG_IMAGES}, we show two examples of DM and stellar haloes with 2D projections of DM density and stellar light, and the corresponding 3D spherically-averaged density profile. These two examples are representative of the relevant mass and radial scales for an elliptical galaxy (upper panels) and a disk-like galaxy (lower panels). For both examples, and for the majority of analyzed haloes, a single power law approximation provides good agreement with the measured stellar halo density profile, from approximately the stellar half-mass radius out to the virial radius.


%
\begin{figure*}
\begin{center}
\includegraphics[width=18cm]{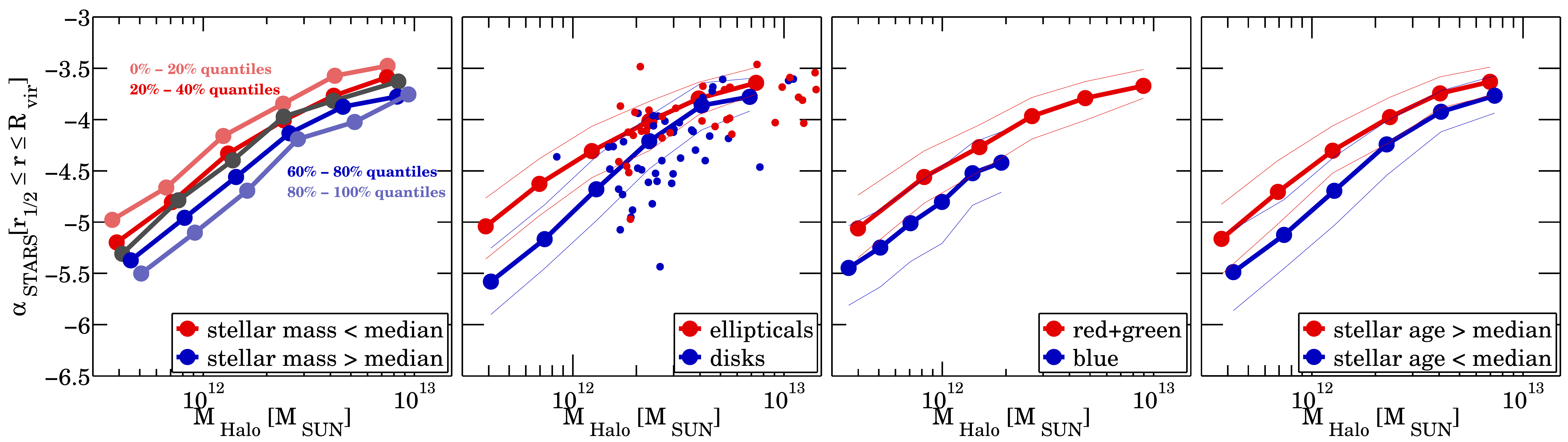}
\caption{The relation between $\AS$ and the properties of the central galaxies, in equally-spaced logarithmic bins of halo mass. Solid thin curves denote, throughout the paper, the 25 and 75 percentiles around the median values (large, filled circles). The blue and red dots of the second panel are the same as in Figure \ref{FIG_STELLARSLOPES}, i.e. hand-picked galaxies with visually inspected morphologies.}
\label{FIG_GALAXYPROPERTIES}
\end{center}
\end{figure*}
%
\section{Relating the stellar halo density profile to halo and galaxy properties}
\label{SEC_MHALO}

In this Section, we study the relationship between the slope of the stellar halo density profile, the properties of the underlying DM haloes, and their central galaxies. 
Figure \ref{FIG_STELLARSLOPES} shows one of the main results of this work: a strong correlation between the slope of the stellar halo and the total mass of the parent dark matter halo, where more massive haloes host shallower stellar haloes (smaller $| \AS |$) than lower-mass haloes. In particular, halo stars fall off much more steeply than the expected outer DM density profiles, with $\AS$ ranging between $-3.5$ and $-5.5$ in the interval $3\times10^{11} \lesssim \MHALO \lesssim 10^{14} \MSUN $. The halo-to-halo variation in stellar halo slope at fixed $\MHALO$ increases towards the low-mass end, reaching $\pm0.5$ (1-$\sigma$) at $\sim10^{12}\MSUN$ and below.

Our resolution tests based on the Illustris simulation suite (see Appendix) indicate that the flattening in the $\AS-\MHALO$ relation below about $3\times10^{11} \MSUN$ is a numerical artifact, marking the lowest mass systems which are sufficiently resolved for this analysis. On the other hand, lack of resolution is not responsible for the enhancement in the scatter towards the low-mass end, which thus appears to be a robust and physical feature. 
Several distinct reasons generate the artificial flattening of the stellar halo slope in poorly resolved haloes: (1) the low-density stellar outskirts are not well sampled when an insufficient number of stellar particles are present, (2) poor resolution enhances tidal stripping of material from orbiting satellites, (3) some level of spurious star formation occurs outside the central galaxy where it is not physically motivated, resulting in too massive {\it in situ} stellar haloes. We find that at least 3000 stellar particles within the virial radius are required to properly characterise the stellar halo structure.

In Figure \ref{FIG_STELLARSLOPES}, green, open symbols represent the same measurements performed on a series of zoom-in cosmological simulations of individual Milky-Way like haloes, characterised by better spatial and mass resolution than Illustris, and implementing a slightly different galaxy formation model and cosmology.
 In particular, square and star symbols represent the Milky-Way like galaxies of \cite{Marinacci:2014}. Based on the same initial conditions of the Aquarius Project haloes \citep{Springel:2008}, they have also been run with {\sc arepo}, at two different resolution levels\footnote{In agreement with the nomenclature of \cite{Springel:2008}, level 5 (4) corresponds to the following values: gravitational softening of $\sim$ 680 (340) pc at $z=0$; $m_{\rm DM} \sim 2 \times 10^6 ~(\sim 3 \times 10^5) ~\MSUN$; $m_{\rm baryon} \sim 4 \times 10^5 (5 \times 10^4)~ \MSUN$.}
 and with a galaxy formation model identical to Illustris' with two exceptions: 1) the stochastically-triggered hot bubbles implemented within the radio-mode AGN feedback have been replaced by a more gentle heating of the halo gaseous atmosphere; 2) the galactic winds implemented within the stellar feedback have been given 50\% of their total energy as thermal energy. Winds are therefore hotter in the \cite{Marinacci:2014} runs than in Illustris when they are launched, helping to prevent spurious in-situ star formation in the circumgalactic medium \citep[see][for more details]{Vogelsberger:2013b, Marinacci:2014}. 

The green triangle represents the structural properties of Eris, a $7\times10^{11} \MSUN$ halo simulated with the N-body+SPH code {\sc gasoline} \citep{Wadsley:2004a} and characterised by a high density threshold for star formation, supernova thermal feedback based on shut-off cooling, and a spatial and DM particle resolution of 120 pc and $4.9\times10^4 \MSUN$, respectively \citep[][]{Guedes:2011, Rashkov:2013}.
Even though Eris' data point is a 1-$\sigma$ outlier, these results demonstrate that the trends identified with the Illustris galaxy population are reliable against numerical limitations, differences in the specific sub-grid physics implementations, as well as changes to the used hydrodynamic algorithm: they confirm that stellar haloes as steep as $\AS^{MW} = -4.5 \pm 0.5$ for Milky-Way like haloes are a robust numerical prediction of our analysis.

As we discuss in Appendix A, the median trend of Figure \ref{FIG_STELLARSLOPES} is insensitive to the specific choice of the radial range over which the stellar halo is fitted with a single power-law formula: the slopes of the stellar haloes as a function of halo mass are consistent, within the 1-sigma variations of Figure \ref{FIG_STELLARSLOPES} and down to the resolution limit, for a variety of choices, and as long as the innermost boundary of the stellar halo is not smaller than $\RHALF$. However, if the slope of the stellar halo is measured over a radial extent which is much smaller than approximately half the virial radius, the halo-to-halo variation is significantly increased due to the occurrence of local overdensities (e.g. shells and streams) as well as profiles better described by a double power-law formula.

Finally, in Figure \ref{FIG_STELLARSLOPES}, red and blue dots represent a sample of hand-picked Illustris galaxies which have been classified as elliptical and disk galaxies by visually inspecting their g, r, i SDSS-band composite light distributions \citep[see Figures 20 and 21 of][]{Vogelsberger:2014b}. While disk-like, star forming galaxies are rare at the highest mass end, their distribution on the $\AS-\MHALO$ diagram at fixed halo mass suggests that the stellar halo slope is connected in some way to the properties of the galaxy residing at its centre. We elaborate on this idea in the next series of plots.

In Figure \ref{FIG_GALAXYPROPERTIES}, we show how the slope of the stellar halo correlates with the properties of the central galaxy: from left to right, stellar masses, morphologies, colors, and stellar population ages. 
Here, large, filled circles denote the median trends in halo mass bins equally separated in logspace, and thus containing variable numbers of galaxies; solid thin curves give the 25 and 75 percentiles of the distributions in the same halo mass bins. 
We define red galaxies as all the simulated systems which lie above the color cut $g-r < 0.4-0.03*(r+20)$, which therefore also includes green-valley galaxies. For all the other panels in Figure \ref{FIG_GALAXYPROPERTIES}, the separation threshold varies according to the halo mass, i.e. according to the distribution of stellar masses, circularities distributions, and stellar ages of the galaxies belonging to the given halo mass bin. Our morphological decomposition is purely kinematic: we classify galaxies based on the fraction of stars with circularities exceeding 0.70, where those lying above the median of all galaxies populating the same halo mass bin are labeled ``disks'', and those below are labeled ``ellipticals''\footnote{Here the circularity of a star is defined as the ratio between its angular momentum component along the disk axis and the angular momentum that the star would have if it were on a circular orbit with the same energy.}. 

Three important conclusions arise from Figure \ref{FIG_GALAXYPROPERTIES}. First, the leftmost panel demonstrates that galaxies which lie above the median trend of the stellar mass -- halo mass relation exhibit steeper stellar haloes (larger $|\AS|$) than less massive galaxies with the same halo mass. Second, at fixed halo mass, disk-like and blue galaxies are surrounded by steeper stellar haloes than elliptical and red galaxies, by up to $\Delta \AS \sim 0.3-0.5$ (two central panels). Finally, galaxies with younger stellar populations (both within the main body of the galaxy and within the stellar halo, rightmost panel) are also characterised by steeper stellar haloes, consistent with them being more disk dominated and bluer than their older counterparts. 
Interestingly, the separation between the different galaxy populations is rather constant as a function of halo mass for all but the morphological classes: the distinction in slope between disk-like and elliptical galaxies vanishes at the high mass end and, more significantly, diverges towards the low-mass end. Furthermore, the halo-to-halo variations in $\AS$ within the same class of objects varies weakly with halo mass, suggesting that the large enhancement in the 1-$\sigma$ variation of Figure \ref{FIG_STELLARSLOPES} is due to physical diversification in the build up of the stellar haloes for different types of galaxies and haloes.
We further comment on this, on the simultaneous relation among $\AS$, $\MHALO$, and $\MS$, and on the implied stellar mass contained in the stellar haloes in the discussion Section.
%

\begin{figure}
\begin{center}
\includegraphics[width=8cm]{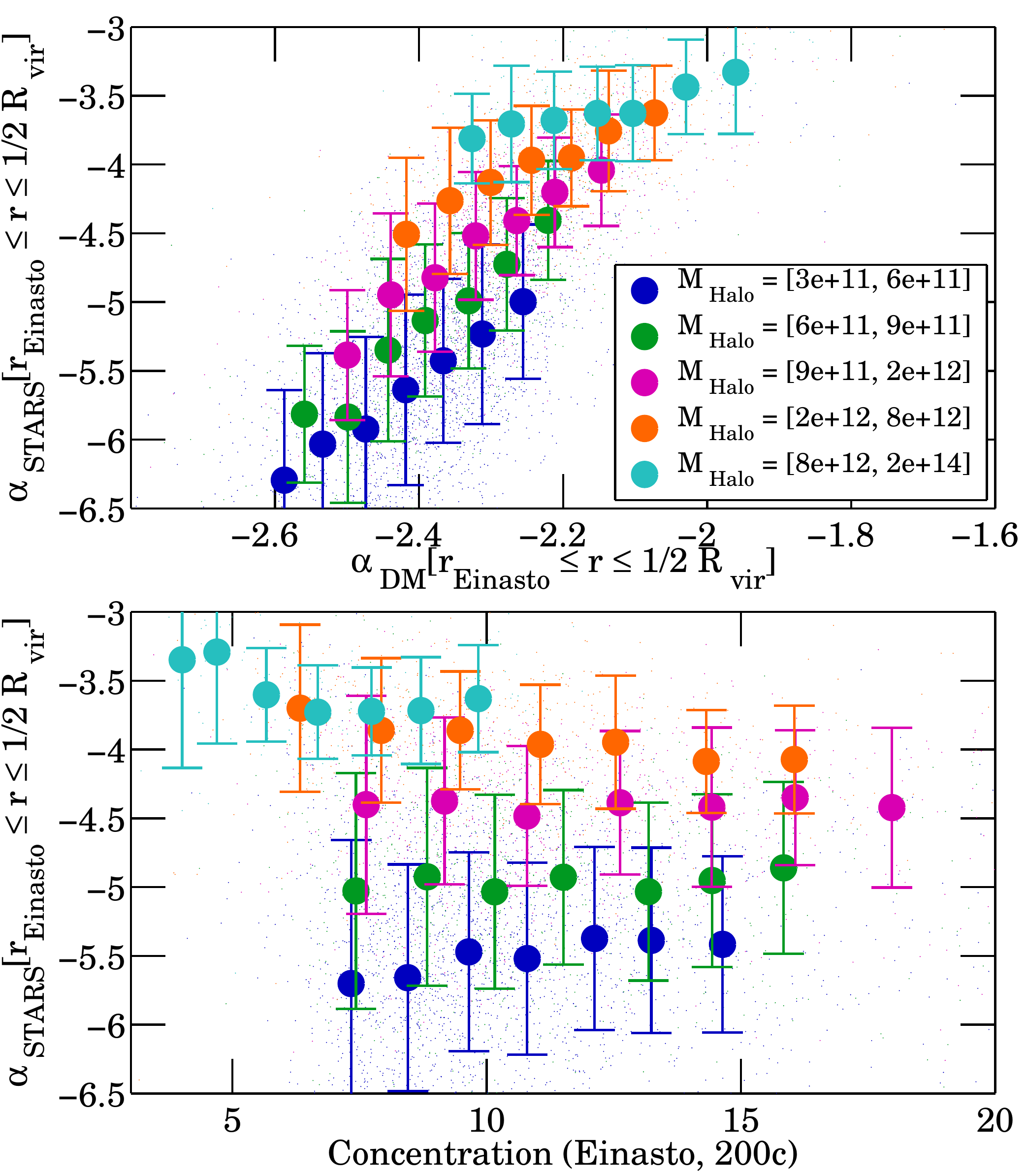}
\caption{Upper panel: the average relation between the slope of the stellar halo ($\alpha_{\rm STARS}$, calculated in the radial range $r_{\rm Einasto} \leq r \leq 1/2 ~\RVIR$) and the slope of the underlying dark matter distribution $\ADM$, measured over the same radial range. 
Here, $r_{\rm Einasto}$ is the scale radius inferred from an Einasto fit of each individual halo, with $r_{\rm Einasto} \gtrsim 2 \times \RHALF $ across the whole halo mass range. The stars are always more centrally concentrated than the underlying DM, which exhibits a lower limit of $\ADM > -2.8$. Lower panel: no clear trend holds between stellar halo slope and DM halo concentration in bins of halo mass; gas cooling and feedback mechanisms in our simulation alter both the concentration--mass and the DM slope--mass relations (see Discussion). Throughout the paper, large, filled circles denote median values, errorbars 1-$\sigma$ standard deviations, in bins of the quantity on the x-axis. Here halo mass in the legend is in units of $\MSUN$.}
\label{FIG_STARSvsDMPROFILES}
\end{center}
\end{figure}

\section{Stars falling away from Dark Matter}
\label{SEC_DM}

To what extent are the halo stars more centrally concentrated than the underlying DM? How does $\AS$ relate to the properties of the underlying DM density profile? 
Can the stellar halo slope provide a constraint for mass modeling, in addition to the broadly used concentration--mass relation seen in N-body only simulations? 
Here we address these questions by quantifying the relationship between these two components.

Although the slope of the logarithmic DM density profiles evolves with distance for pure NFW or Einasto haloes, the same profiles can still be well-fitted by a single power law over a restricted radial range within the virial radius. Moreover, gas cooling and feedback can significantly alter the DM profiles, sufficiently so that the NFW and Einasto models become poor descriptions of the haloes: for example, we have noticed that when contraction occurs, DM profiles immediately outwards of the scale radius exhibit reduced curvature compared to their N-body only analogs \citep[see][for the preliminary analysis of the effects of baryons on the underlying DM structures in Illustris]{Vogelsberger:2014, Genel:2014}.

To minimize this problem, we have fit each individual halo with a single power law in the range $r_{\rm Einasto} \leq r \leq 1/2 ~\RVIR$, where $r_{\rm Einasto}$ is the scale radius obtained via Einasto fitting, which is at least two times larger than the stellar half-mass radius (see Appendix for details). 
We measure the logarithmic slopes of both stars and DM in this same radial range. 
The upper panel of Figure \ref{FIG_STARSvsDMPROFILES} shows that the DM halo slopes beyond the scale radius are effectively confined between $-2.8$ and $-1.8$ at all analyzed halo masses. The DM densities fall off steeper beyond the scale radius for increasingly smaller haloes. Importantly, they are remarkably well correlated with the stellar halo slopes, both at fixed halo mass and across masses (although the relation becomes flatter for the largest masses). However, as discussed in the previous Section and confirmed here across a different radial range, the stellar haloes span a much larger range of logarithmic slope values, namely: $-7 \lesssim \AS \lesssim -3$. The stellar haloes are thus much steeper than the underlying DM. 
This conclusion is again insensitive to the exact choice of the inner radius for the comparison between stars and DM, as long as this radius is chosen to be beyond about twice the stellar half-mass radius. Similarly, moving the outer boundary out to the virial radius only reduces the values of the DM slope by $\Delta \ADM \sim 0.2$.

%

Prompted by the strong average correlation between stellar and DM slope, we investigate the relation between stellar halo slope and DM halo concentration in the lower panel of Figure \ref{FIG_STARSvsDMPROFILES}. Here we report the concentrations derived via Einasto fitting, but the same conclusions hold for NFW concentrations and with alternative non-parametric methods of measuring the scale radius. 
There is no clear trend between $\AS$ and concentration at fixed halo mass, while we would have expected the steepest stellar haloes to be associated with the most concentrated DM haloes. 
We comment further on this null result in the following sections.



\begin{figure*}
\begin{center}
\includegraphics[width=17cm]{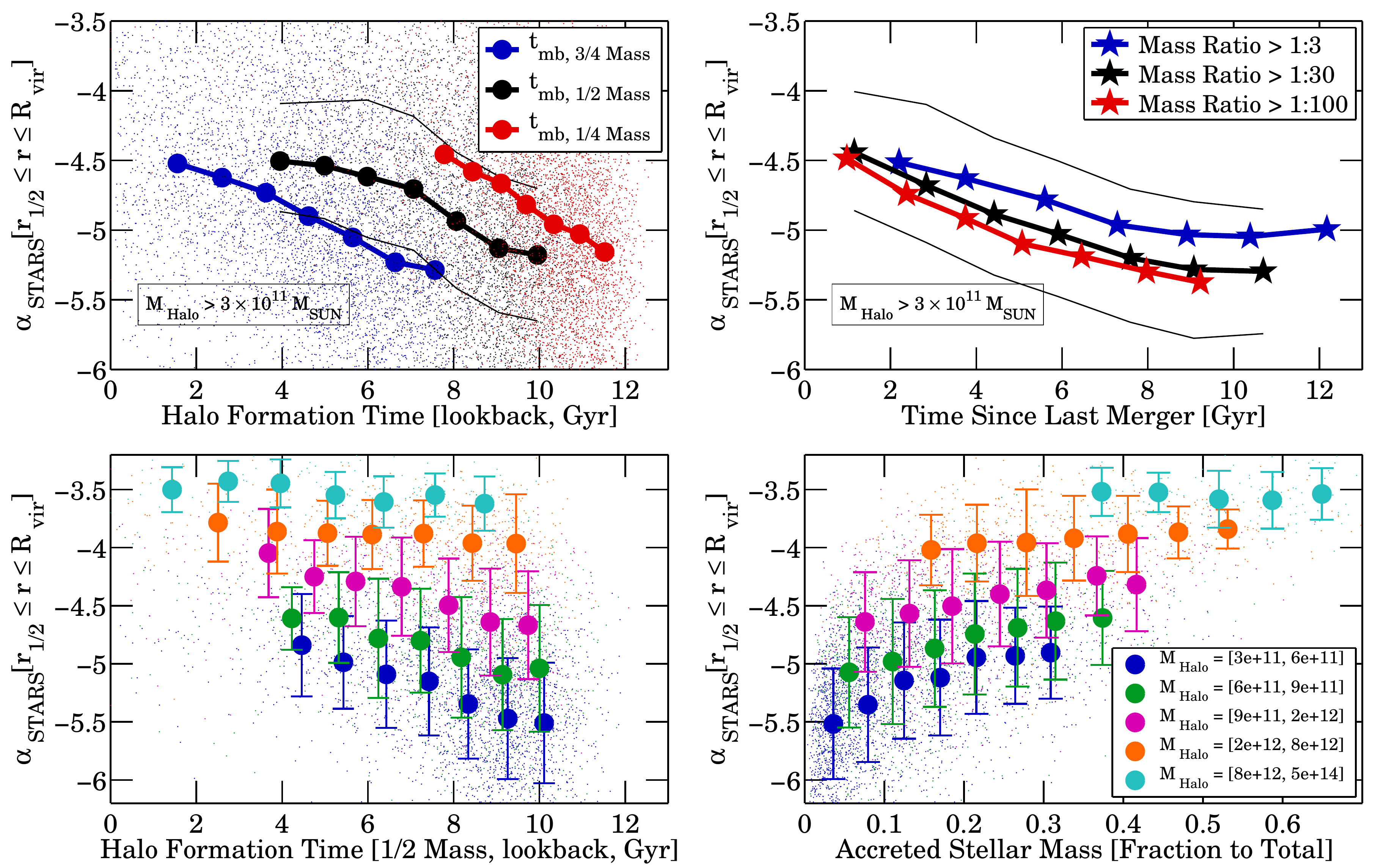}
\caption{Halo accretion history imprinted in the stellar halo slope. Upper panels: stellar halo slope as a function of halo formation time (top left) and the time since the last merger (top right), for all haloes above the resolution threshold. 
In the top left panel, the red, black, and blue colors refer to early ($t_{\rm mb, 1/4 Mass}$), intermediate ($t_{\rm mb, 1/2 Mass}$), and recent ($t_{\rm mb, 3/4 Mass}$) formation times, i.e. to the times at which one quarter, half or three quarters of the total mass at $z=0$ have been assembled in a halo along its main progenitor branch, respectively.
We observe a general trend that haloes formed more recently have shallower stellar haloes. 
Bottom left panel: the relation between $\AS$ and the intermediate formation time ($t_{\rm mb, 1/2 Mass}$) is preserved at fixed halo mass towards the low mass end, being the trend progressively stronger for halo masses smaller than about $10^{13}\MSUN$. 
Bottom right panel: dependence of $\AS$ on the fraction of stars accreted from infalling satellites and mergers as a function of halo mass (halo mass in the legend is in units of $\MSUN$).}
\label{FIG_HISTORY}
\end{center}
\end{figure*}

\section{Halo Accretion History imprinted in the Stellar Halo Slope}
\label{SEC_ASSEMBLYHISTORY}

In Section \ref{SEC_MHALO} we motivated the scatter in the $\AS-\MHALO$ relation by distinguishing among the properties of the central galaxies. Here we search for the origin of such variations by characterising the assembly history of the DM haloes. Inspired by the results of \cite{Deason:2013} and \cite{Deason:2014a}, obtained via the hybrid semianalytic+N-body simulations applied to 12 Milky-Way like galaxies \citep{Bullock:2005}, we demonstrate and quantify that the slope of the stellar halo density encodes a wealth of information about the assembly history of the haloes, free from selection biases, across a wide range of masses. 

In Figure \ref{FIG_HISTORY}, we show that DM haloes which formed more recently (left column), which experienced a more recent major/minor merger (top right) or that accreted larger fractions of stellar mass from infalling satellites (bottom right) exhibit shallower stellar haloes than their older counterparts by up to $\Delta \AS \sim 0.5-0.7$ at fixed halo mass. 
In particular, in the top panels of Figure \ref{FIG_HISTORY}, we show $\AS$ as a function of formation time for all halo masses above our resolution threshold, and for six different definitions of halo age: 1) the time at which one quarter, half or three quarters of the total mass at $z=0$ have been assembled in a halo along the progenitor branch which maximizes the cumulative mass (this is the main branch of a halo, to which the subscript ``mb'' refers to in $t_{\rm mb, 1/4 Mass}$, $t_{\rm mb, 1/2 Mass}$, and $t_{\rm mb, 3/4 Mass}$, respectively: top left panel); 2) the time of the last merger, defined to have stellar mass ratio at infall larger than 1:3, 1:30, and 1:100 (top right panel).
More massive haloes have both shallower stellar slopes, and more recent formation times, than smaller haloes, where $t_{\rm mb, 1/2 Mass}= $ 5 Gyrs ago on average at $\sim10^{14}\MSUN$, compared to about 9 Gyrs ago for $3\times 10^{11}\MSUN$ haloes, although large halo-to-halo variations exist.

The values that $\AS$ spans as a function of halo age in Figure \ref{FIG_HISTORY} (upper panels) appears too small to justify the range from Figure \ref{FIG_STELLARSLOPES} across the same halo mass interval; this is simply because the median trends (large, filled circles) are given in bins of halo formation time, and so are necessarily lowered by the low-mass halo population. In fact, it is important to point out that the effect of the halo age on the steepness of the stellar halo is also preserved at fixed halo mass, even though the trend is flat for halo masses larger than about $10^{13}\MSUN$ and becomes progressively steeper for increasingly smaller DM haloes (bottom left panel): we expand and capitalize on this in the next Subsection.

Interestingly, different parameterisations of the halo age correlate more or less strongly with the slope of the stellar halo: in the top right panel, we show that the time of the last {\it major} merger is a worse proxy for the stellar slope than the halo formation redshift of the top left panel, regardless of the definition. We also find that whether the last merger is major (1:3) or minor (1:10) is rather insignificant (not shown). However, the time of the most recent accretion of a satellite with smaller mass ratio appears more relevant for the build-up of the stellar halo: requiring that a halo not have accreted any satellite with mass ratio larger than 1:100 in the last e.g. 10 Gyrs is indeed a much stronger constraint on the quiescence of its assembly history than requiring it not have merged in the same span of time with a galaxy with mass ratio larger than 1:3.
%
By simultaneously decomposing the dependence of $\AS$ on halo mass and halo age, we also find that parameterisations of the halo age which are more sensitive to the recent assembly history ($t_{\rm mb, 3/4 Mass}$) than to the early halo assembly ($t_{\rm mb, 1/4 Mass}$) show stronger trends with $\AS$ for masses below $10^{12}\MSUN$, although this is not explicitly captured in Figure \ref{FIG_HISTORY}.

Finally, in the bottom right panel of Figure \ref{FIG_HISTORY}, we consider one last quantity that encodes the essence of the hierarchical growth of structures in CDM scenarios -- namely, the fraction of stellar mass that a galaxy (main body + stellar halo) accreted from mergers as well as disrupted satellites.
We define ``ex-situ'' or ``accreted'' as those stars that, at the time of formation, were bound to any halo which lie outside the main progenitor branch of a given halo at $z=0$.
Given this definition, the ex-situ fraction grows rapidly with halo mass, up to $\sim$70--80 percent for galaxies at the centres of the clusters.
Larger fractions of ex-situ stars also induce shallower stellar haloes, but this trend flattens for halo masses larger than about $10^{13} \MSUN$. 
For haloes below this mass, our analysis suggests that elliptical galaxies which are hosted by more recently formed DM haloes exhibit larger accreted stellar mass fractions (see \textcolor{blue}{Rodriguez-Gomez \etal 2014b, in preparation}), and consequently have shallower stellar halo profiles.

\begin{figure}
\begin{center}
\includegraphics[width=8.5cm]{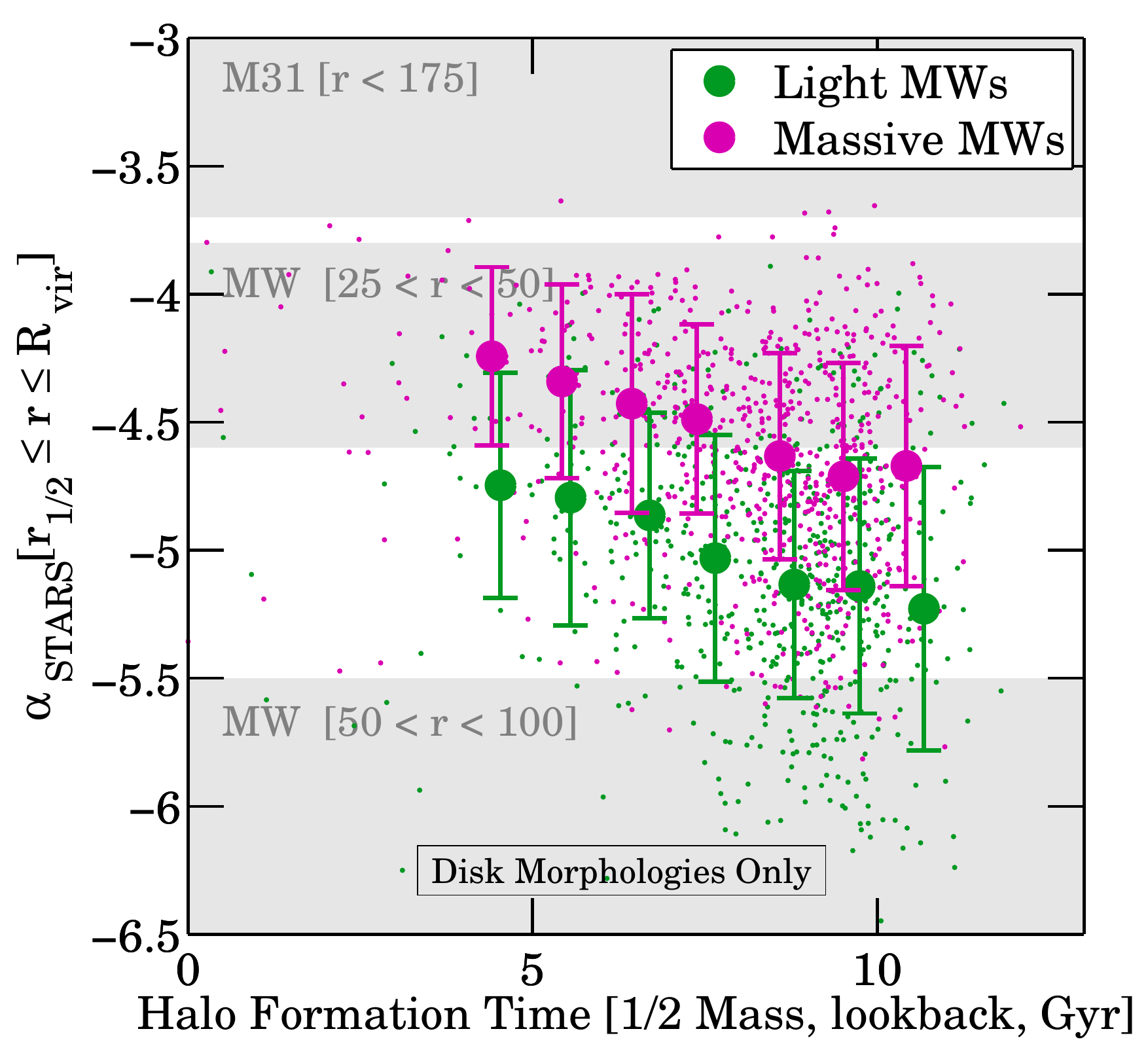}
\caption{The relation between the stellar halo slopes and the halo formation time for Milky-Way-like galaxies (light: $6 \times 10^{11} < \MHALO \le 9 \times 10^{11} \MSUN$; massive: $9 \times 10^{11} < \MHALO \le 2 \times 10^{12} \MSUN$). The trend between slope and formation time is preserved also at fixed halo masses, and becomes steeper for smaller masses. Here, in contrast to Figure \ref{FIG_HISTORY}, only galaxies exhibiting disk-like morphologies are considered: note that, for Milky-Way masses, the average halo formation times for disk galaxies are larger by $\sim 0.6-0.7$ Gyrs than for the whole population. The grey bands denote current observational constraints for M31 and the MW, performed across different radial ranges (see text for details).}
\label{FIG_MWs}
\end{center}
\end{figure}
\subsection{Galactic Archeology with the Stellar Halo Slope}

In order to explore the implications of our findings for the history of the Local Group, we now focus exclusively on haloes with Milky-Way like masses.
We have 1744 such galaxies in Illustris satisfying our selection criteria (see Section \ref{SEC_SAMPLE}), half of which we define to have disk-like kinematics, and which we separate in two halo mass bins, namely $6 \times 10^{11} < \MHALO \le 9 \times 10^{11} \MSUN$ for ``light'' Milky Ways, and $9 \times 10^{11} < \MHALO \le 2 \times 10^{12} \MSUN$ for ``massive'' Milky Ways \citep[see][for a detailed discussion about the constraints on the Milky Way mass]{Watkins:2010}. 
As stated in Section \ref{SEC_MHALO}, we find that Milky-Way like haloes (combining light+massive) have $\AS^{MW} \sim -4.5$, lowered by an additional $0.2-0.3$ when considering only disk galaxies. 

Figure \ref{FIG_MWs} quantifies the dependence of the stellar halo slopes on the combination of DM halo mass, galaxy morphology, and assembly history. Here we demonstrate that at fixed mass and fixed morphology, a residual trend with halo formation time is manifest: galaxies that assembled half of their total halo mass within the last 5 Gyrs have stellar profiles shallower by $\simeq 0.5$ than haloes with formation times more than 10 Gyrs ago. 
Moreover, we find that, according to our definition, disk and elliptical galaxies reside in haloes whose median halo formation time differs by up to $\sim 1.2$ Gyrs at fixed halo mass, although with similarly large 1-$\sigma$ variations within each morphological class \citep[a result which will be expanded elsewhere; see also][]{Sales:2012}.
We note that at these masses, the difference between a parameterisation based on $t_{\rm mb, 1/2 Mass}$ or $t_{\rm mb, 3/4 Mass}$ is negligible.
However, at fixed halo mass, the distribution of the halo formation times is very asymmetric, peaking e.g. at 8.9 Gyrs ago for light MWs and being skewed towards more recent formation times with an extended tail: the smallest values of $|\AS|$ are indeed recorded in such tails of the halo formation time distribution.
Finally, no strong residual dependence can be found as a function of accreted stellar fraction or as a function of the total number of accreted satellites, once halo mass, morphology and halo formation time are fixed.

A comparison of our simulated results with current observational constraints of the stellar halo slopes for M31 and the Milky Way is potentially enlightening: these are given in Figure \ref{FIG_MWs} by grey areas.
While the stellar halo of M31 is remarkably shallow across a wide range of radial distances and measurements \citep[$-3.7 \lesssim \AS^{M31} \lesssim -3$: ][]{Gilbert:2012, Ibata:2014}, for the Milky Way, current assessments based on stellar number counts depend sensitively on the galactocentric distance probed by the measurements. Within $r \sim 20-30$ kpc, \cite{Bell:2008, Watkins:2009, Sesar:2011, Deason:2011a} find a power law slope of $-3 \lesssim \AS^{MW} \lesssim -2$, but that a steeper index is required for distances between 25 and 50 kpc, namely $-4.6 \lesssim \AS^{MW} \lesssim -3.8$. Recently, \cite{Deason:2014a} found that A-type stars selected from the Sloan Digital Sky Survey at $r \sim 50-100$ kpc are well modeled by an even smaller index, $\AS^{MW} = -6 \pm 0.5$. 

It is important to emphasise that the trend we find between stellar halo slope and halo formation time is qualitatively confirmed regardless of the choice of the radial range over which the simulated stellar haloes are considered, but with a caveat: measurements over smaller radial extents result in progressively noisier and weaker relations, with larger halo-to-halo variations at fixed formation time. Overall, we find that stellar haloes characterised across a radial extent which is larger than approximately half the virial radius are less sensitive to local inhomegenities, and are thereby more stable proxies of the global properties of the underlying DM haloes and central galaxies. 
In this respect, it is plausible that observational measurements performed simultaneously across the {\it whole} available galactocentric distances within the Milky Way would best agree with the locus identified by our simulation results for either light or massive Milky Ways.

Remarkably, our simulation results seem to be more consistent with the relatively steep stellar halo measured for the Milky Way, and broadly inconsistent with the shallow values measured for M31. 
The discrepancy between the halo slopes of Milky Way and M31 can thereby be interpreted with a possibly large diversity between the two galaxies in halo mass, assembly history, or a combination of both.
In the case of the Milky Way, stellar slopes as small as $-6$ can result in haloes with formation times as large as $7-10$ Gyrs ago, consistent with a series of considerations which support a much more active (recent) assembly history for M31 than for the Milky Way: M31 has a more disturbed disk \citep{Brown:2006}, a larger bulge \citep{Pritchet:1994, Durrell:2004}, a younger and more metal-rich stellar halo population \citep{Irwin:2005, Kalirai:2006}, more numerous tidal streams and satellites \citep{Koch:2008, Richardson:2011}, and more numerous and more massive globular clusters \citep{Huxor:2014} than the Milky Way.
On the other hand, our theoretical data seem to favor a large mass for M31 ($M_{\rm 200c} \gtrsim 2 \times 10^{12} \MSUN$), adding yet a cautious voice to the highly debated and uncertain issue of the Local Group mass \citep[see][for a summary and the most recent arguments, respectively]{Watkins:2010, Penarrubia:2014v2, Diaz:2014}.

\section{Discussion}
\label{SEC_DISCUSSION}

Our numerical results confirm the cosmological origin of stellar haloes, whose smooth and less smooth density distributions are the result of stellar stripping from accreted and merging satellites.
%
The steepening of the stellar halo slope towards the low-end of the halo mass function (Figure \ref{FIG_STELLARSLOPES}) can be interpreted as the quintessential manifestation of the hierarchical growth of structure in CDM scenarios {\it and} the way galaxies assemble at the centres of the DM potential wells.
%
More massive haloes accrete more numerous and more luminous satellites than their low-mass companions \citep[e.g.][]{Gao:2004}, and stochasticity in the star-formation is expected to have larger impact at progressively smaller halo masses \citep[e.g.][]{Kuhlen:2013, Sawala:2013, Shen:2013b, Sawala:2014}. 
Moreover, more massive haloes accrete the majority of their mass only recently, 
and recently accreted satellites tend to have larger apocentres than if they were accreted at earlier times when haloes were smaller \citep{Gao:2004, Cooper:2010, Rocha:2012}, resulting in less centrally-concentrated stripped material.
However, at fixed amount of accreted stellar mass or at fixed halo mass, the organization of such material across the halo volume further depends on the halo formation time of the host, and on the galaxy type. 
%
Numerical, controlled experiments seem to support the idea that the presence of a disk can enhance the disruption of satellite galaxies in the inner parts of haloes, possibly because of disk shocking or enhanced tidal encounters \citep{Donghia:2010, Penarrubia:2010}.

In Section \ref{SEC_DM}, we compared the stellar density profiles to the underlying DM's, and confirmed that stars fall off much more steeply than the DM \citep[see e.g.][Figure 3]{Abadi:2006}. Indeed, it is expected that halo stars are more centrally concentrated than the dark matter. Stars populate the stellar halo exclusively because they get stripped from infalling satellite galaxies \citep[e.g.][]{Bullock:2005} -- or because they are heated up from the disk or innermost regions of a galaxy \citep{Zolotov:2009, Purcell:2010}, although in minor fractions \citep{Pillepich:2014c}. On the other hand, the DM that populates halo outskirts can accrete either smoothly or as part of infalling substructures, which can then be subsequently stripped because of tides. Due to differences in binding energy, DM and stars are stripped differently and at various times along the orbital history of a subhalo/satellite. The impact of stripping also differs due to the fact that not all subhaloes host a galaxy, driven by stochasticity in star formation, especially at the low-mass end.
The novelty of our contribution consists in having quantified -- for the first time across a large range of halo mass, and with a self-consistent model for the formation of diverse galactic morphologies within the full cosmological context -- that the power law indices of the 3D stellar and DM density profiles in the outskirts of galaxies can differ by up to $\Delta |\ADM - \AS| \sim 4$ for $\sim 10^{11} \MSUN$ haloes. 

Finally, as for the absence of a strong, clear relation between stellar halo slopes and DM halo concentrations, we think that a combination of factors is the culprit. 
First, within our implementations of the subgrid physics, DM haloes respond to baryonic effects in a non-monotonic mass-dependent way \citep{Genel:2014}, e.g. in the mean mass--concentration relation. Moreover, baryonic physics largely enhance the scatter both in the aforementioned relation as well as in the up-to-now unexplored relation between DM slopes and halo masses.
Secondly, DM haloes are more sensitive to the {\it early} assembly of their material (e.g. $t_{\rm mb, 1/4 Mass}$), while stellar haloes seem to be more dependent on the late, more recent formation time (e.g. $t_{\rm mb, 3/4 Mass}$, at least for low-mass objects).
As a consequence, it is plausible that other implementations of the sub-grid physics might result in a qualitatively different conclusion.

\subsection{A new ruler to infer the Halo Mass?}
\label{SEC_SHARELATION}
Phenomenologically, the variation in the stellar halo slope as a function of halo mass can be parameterised by accounting for the stellar mass of the central galaxy: galaxies which lie below the median trend of the stellar mass -- halo mass relation exhibit shallower stellar haloes than more massive galaxies with the same halo mass
(see Figure \ref{FIG_GALAXYPROPERTIES}). 
Conversely, the simultaneous relation among $\AS$, $\MS$, and $\MHALO$ might provide, in principle, a new, potentially-powerful tool to infer halo masses via the measurement of the stellar halo slope and the stellar mass of the central galaxies.
We find that the uncertainties in the inferred halo masses from halo abundance matching alone can be improved by an additional $20-30\%$ when constraints from the stellar halo slopes measured at reasonable accuracy are added. 
We will further explore this possibility in future work, by better quantifying the errors due to the convolution of all the distinct and yet correlated effects mentioned above.
%
While we appreciate this exciting possibility in light of current and future measurements of the low-surface brightness features in distant galaxies \citep[e.g. with the Dragonfly Telephoto Array:][]{Abraham:2014, Dokkum:2014, Zackrisson:2012}, we are nevertheless cautious of the perils and the difficulties of calibrations based exclusively on theoretical models, however self-consistent.

\subsection{Mass enclosed in the Stellar Halo}

\begin{figure}
\begin{center}
\includegraphics[width=8.5cm]{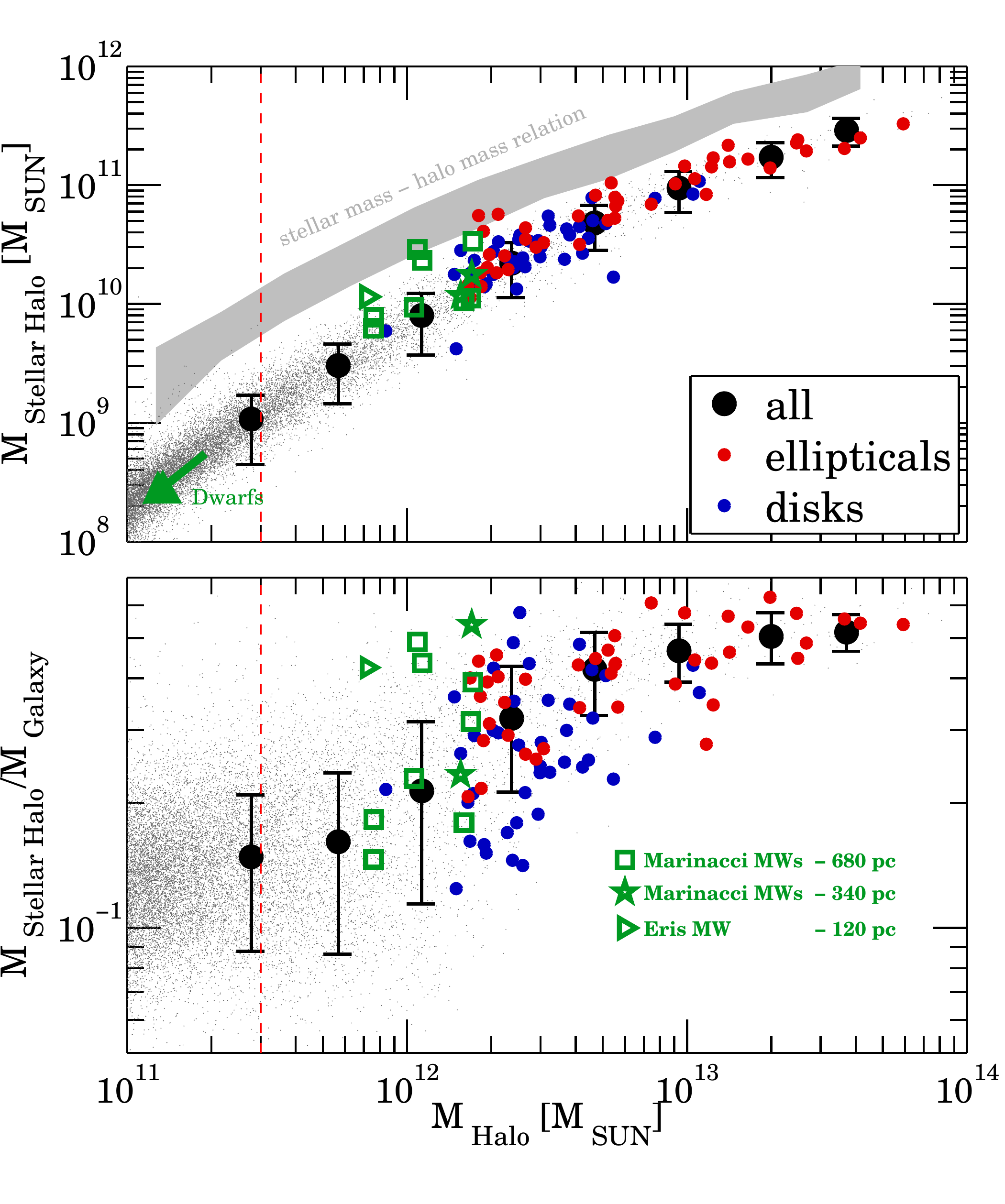}
\caption{Upper panel: stellar mass enclosed in the stellar halo ($r > 2\times r_{1/2}$) as a function of halo mass. Lower panel: ratio of the stellar halo mass to the stellar mass of the central galaxy ($r < 2\times r_{1/2}$). Open, green symbols represent a sample of individually-simulated zoom-in galaxies, as in Figure \ref{FIG_STELLARSLOPES}. The gray band in the upper panel identifies the locus of the stellar mass -- halo mass relation in Illustris. The arrow in the lower left corner refers to the outcome of two dwarf galaxies simulated at $\sim$60-80 pc resolution \citep[][see text for details]{Vogelsberger:2014c}. It is important to keep in mind that the actual definition of stellar halo mass is {\it arbitrary}: given the choice for the boundary that distinguishes between main galaxy and stellar halo, the same numerical results at different resolution appear to be more dependent on the physical extent of the galaxy body rather than a modification to the mass content. For Eris, for example, a different separation, one that better mimics a photometric morphological decomposition, results in a stellar halo whose mass amounts to 7\% of the galaxy mass, instead of the $\sim$ 40\% value reported here \citep[see][]{Pillepich:2014c}. 
}
\label{FIG_STELLARHALOMASS}
\end{center}
\end{figure}

The acquisition of a future large sample of galaxies with radial profiles reaching very low surface densities will not only aid the calibration of the $\AS-\MHALO$ relationship, but also the inferred relation between mass enclosed in the stellar halo and mass of the central galaxy. We show these in Figure \ref{FIG_STELLARHALOMASS}, where the mass of the stellar halo (of the galaxy) is {\it arbitrarily} defined as all material beyond (within) $2\times r_{1/2}$, down to the virial radius.
In relation to the functional form chosen by \cite{Moster:2013} for the stellar mass--halo mass connection, our data is well reproduced by the following fitting formula

\[
m_{*} = 2 m_{\rm h} N \left[ \left(\frac{ m_{\rm h}}{M_1}\right)^{-\beta} + \left(\frac{ m_{\rm h}}{M_1}\right)^{\gamma} \right]^{-1}
\]
where $m_{\rm h} \equiv {\rm log}_{10} ~\MHALO$ and $ m_{*} \equiv {\rm log}_{10} ~\MSH$, and the best-fit parameters read $[M_1, N, \beta, \gamma] = [13.74, 0.8432, 2.434, 2.999]$ for the median trend and [13.82, 0.7967, 2.191, 2.979] for the 1-$\sigma$ scatter, respectively.\footnote{In light of the strong, quantitative dependence of $\MSH$ on the exact definition of what the stellar halo is and what its radial boundaries are, we encourage the interested readers to get in contact with the authors for measurements of the Illustris stellar halo masses which are better optimized to the individual needs.}

The comparison of our findings to the outcome of zoom-in, N-body+hydrodynamics simulations of Milky-Way like galaxies (green, open symbols, as in Figure \ref{FIG_STELLARSLOPES}) is encouraging, and results once more in both an additional resolution convergence test as well as a check on the possible influence of the specific sub-grid choices implemented in Illustris. 
The red and blue dots in Figure \ref{FIG_STELLARHALOMASS}, which represent visually-inspected elliptical and disk-like galaxies respectively, hint in the lower panel towards a dependence of the stellar halo mass fraction on morphology, in agreement with what is found for the stellar halo slope. 

For Milky Way like galaxies, within our definition, the stellar halo mass can vary considerably, from 10 to 30\% of the galaxy stellar mass. This might appear at odds with the few, currently available observational inferences of the mass enclosed in the stellar haloes of the Galaxy, M31, NGC 351 and M101 \citep{Carollo:2010, Courteau:2011, Ibata:2014, Bailin:2011, Dokkum:2014}, which all but perhaps for M31 seem to favor mass fractions of the order of a few percents and below.
However, the choice for the boundary that distinguishes between main galaxy and stellar halo is crucial, and varying the adopted stellar halo edges (both inner and outer) can change the stellar halo mass fractions significantly. Thereby, great care must be taken when comparing the median trends presented here to observational constraints which have not been measured in a self-consistent way. 
In contrast, the slope of stellar halo which has been the main focus of this paper is a much more robustly-defined and informative observable, and differently for the stellar profile normalization -- which is a more direct proxy of $\MSH$-- it is less sensitive to the star-formation efficiency within the accreted former satellite galaxies: we thereby advocate $\AS$ should be preferred to characterize the stellar halo.

%

\subsection{Outlook}

From the theoretical point of view, the results presented here will need to be better calibrated also with the aid of zoom-in simulations across all halo masses, as for example with the suite of individually-simulated high-resolution galaxies of the MaGICC sample, the AGORA project, and the FIRE simulations by, respectively, \cite{Stinson:2012, Kim:2014, Hopkins:2013}. 
In the meantime, it is interesting to notice that the amount of mass enclosed in the stellar halo of two dwarf galaxies simulated at about 60-80 pc resolution by \cite{Vogelsberger:2014c} falls within the extrapolation of the Illustris data point, with $\rm M_{stellar halo} = 2.5\times 10^7 \MSUN$ and $1.9\times 10^8 \MSUN$ for $\MHALO = 1.2\times 10^{10} \MSUN$ and $7.1\times 10^{10} \MSUN$, respectively (see arrow in the upper panel of Figure \ref{FIG_STELLARHALOMASS}).

At the other end of the halo mass function, our results in Figure \ref{FIG_STELLARHALOMASS} exhibit a flattening of the stellar halo mass ratio to the BCG mass, at about 50--60 per cent. This confirms the findings of \cite{Purcell:2007}, who via semi-analyitical models already found a similar trend between intrahalo light and halo mass beyond the group scale, although with a plateau a factor of 2 or 3 lower. Such flattening pairs with the analogous behavior of the $\AS-\MHALO$ relation at masses larger than $\sim 3\times 10^{13}\MSUN$, with a corresponding decrease in the halo-to-halo variation. 
Indeed, all the relationships found in this work between stellar halo slope and DM slope (Figure \ref{FIG_STARSvsDMPROFILES}), halo formation times (Figure \ref{FIG_HISTORY}, lower left panel) and stellar ex-situ fraction (Figure \ref{FIG_HISTORY}, lower right panel) become progressively flatter or weaker for increasingly more massive haloes, whose assembly is too recent to have allowed a diversification in the build-up of their stellar haloes/ICL. 
Our simulated data thereby predicts that $10^{15}\MSUN$ clusters should be surrounded by ICL with power-law indexes fixed at about $-3.5$, averaged across their entire haloes.

We defer the task of quantifying such predictions in more observationally-oriented terms to future works: these will include the measurement of the 2D stellar halo slopes from the synthetic surface brightness profiles of Illustris galaxies at $z=0$ and at higher redshift (for the outer galaxy populations), and the assessment of the variations in the measurement of the stellar slope of our Milky Way due to the selection of limited fields of view and specific sub-populations of stars. 

\section{Summary and Conclusions}
\label{SEC_SUMMARY}

In this paper, we have uncovered a series of results regarding the properties of the stellar haloes and their relation to the underlying DM haloes, their central galaxies, and their halo assembly histories. To do so we have used Illustris, a state-of-the-art simulation which combines the statistical power of a $\sim$106 Mpc-side cosmological volume with gasdynamics, prescriptions for star formation, feedback, and kpc resolution. 
In particular, we have analyzed a sample of about 5,000 well-resolved galaxies at $z=0$, and measured the power-law index of the smooth, 3D, spherically-averaged density profiles of both stars and DM in their outer regions ($r_{1/2} < r \leq \RVIR$).
Our main findings can be summarized as follows:

\begin{itemize}

\item A strong trend holds between the slope of the stellar halo and the total mass of the parent DM halo: more massive DM haloes host shallower stellar haloes (smaller $|\AS|$) than lower-mass counterparts, with $-5.5 \pm 0.5 < \AS < -3.5 \pm 0.2$ in the studied mass range $3\times 10^{11} \lesssim M_{\rm halo} \lesssim 1\times 10^{14} \MSUN$ (Figure \ref{FIG_STELLARSLOPES}).\\

\item The amount of scatter in the $\AS$--$\MHALO$ relation increases towards the low-mass end, and it is due to the physical diversification in the build up of the stellar haloes for galaxies of different types and haloes of different {\it ages}.\\

\item At fixed halo mass, elliptical, red and old galaxies are surrounded by shallower stellar haloes than disk-like, blue, younger galaxies (Figure \ref{FIG_GALAXYPROPERTIES}).\\

\item DM haloes which formed more recently, which experienced more recent accretion and merger events, or that accreted larger fractions of stellar mass from infalling satellites exhibit shallower stellar haloes than their older analogs, by up to $\Delta \AS \sim 0.5-0.7$, for given halo mass and galaxy type (Figures \ref{FIG_HISTORY} and \ref{FIG_MWs}).\\

\item The slope of the stellar halo density profile is much steeper than the underlying DM density, by up to $\Delta |\ADM - \AS| \sim 4$ for $\sim 10^{11} \MSUN$ haloes, immediately beyond the scale radius. At fixed halo mass, steeper stellar haloes are associated to steeper DM density profiles; however, no clear trend can be found in Illustris between the stellar halo slope and the DM halo concentration in given halo mass bins (Figure \ref{FIG_STARSvsDMPROFILES}) .\\

\item For Milky-Way like galaxies ($6\times 10^{11} \lesssim M_{\rm halo} \lesssim 2\times 10^{12} \MSUN$), we find typical values of the stellar halo slope of $\AS = -4.5 \pm 0.5$, consistent with the results from zoom-in, N-body+hydrodynamics simulations characterised by higher resolution and alternative sub-grid physics prescriptions (Figure \ref{FIG_STELLARSLOPES}). \\

\item In light of our numerical findings, the strikingly different measurements of the outer stellar slopes in the Milky Way and M31 seem to favor a massive M31 ($\MHALO \gtrsim 2 \times 10^{12} \MSUN$), and a Milky Way featuring a much quieter accretion activity in the last 10 Gyrs than its companion (Figure \ref{FIG_MWs}).
\end{itemize}

The shape and amount of scatter of the $\AS$--$\MHALO$ relation quantified in this work could not have been obtained without a self-consistent model for the formation of an acceptable mix of galactic morphologies in a full cosmological context. 
The results summarized here, together with the intriguing perspective of a useful threefold relation among halo mass, stellar mass, and stellar halo slope (see Figure \ref{FIG_GALAXYPROPERTIES} and Section \ref{SEC_SHARELATION}), hopefully will aid the interpretation and foster the advent of deeper surface brightness data of larger samples of galaxies \citep[e.g. with HST or the Dragonfly Telephoto Array:][]{Abraham:2014, Dokkum:2014}.

\section*{Acknowledgements}
AP thanks Akos Bogdan, Arjun Dey, Tomer Tal, and Vasily Belokurov, for useful discussions. FM acknowledges
support by the DFG Research Centre SFB-881 'The Milky Way System' through project A1. AP, AD, LS thank the Aspen Center for Physics and the NSF Grant \#1066293 for hospitality during the final editing of this paper.
VS acknowledges support by the European Research Council under ERC-StG grant EXAGAL-308037.  LH acknowledges support from
NASA grant NNX12AC67G and NSF award AST-1312095.
Simulations were run on the Harvard Odyssey and CfA/ITC clusters, the Ranger and Stampede supercomputers at the Texas Advanced Computing Center as part of XSEDE, the Kraken supercomputer at Oak Ridge National Laboratory as part of XSEDE, the CURIE supercomputer at CEA/France as part of PRACE project RA0844, and the SuperMUC computer at the Leibniz Computing Centre, Germany, as part of project pr85je.

\appendix
\label{SEC_APPA}

\section{Sample, Radial Ranges and Convergence Tests}

\begin{figure}
\begin{center}
\includegraphics[width=8.5cm]{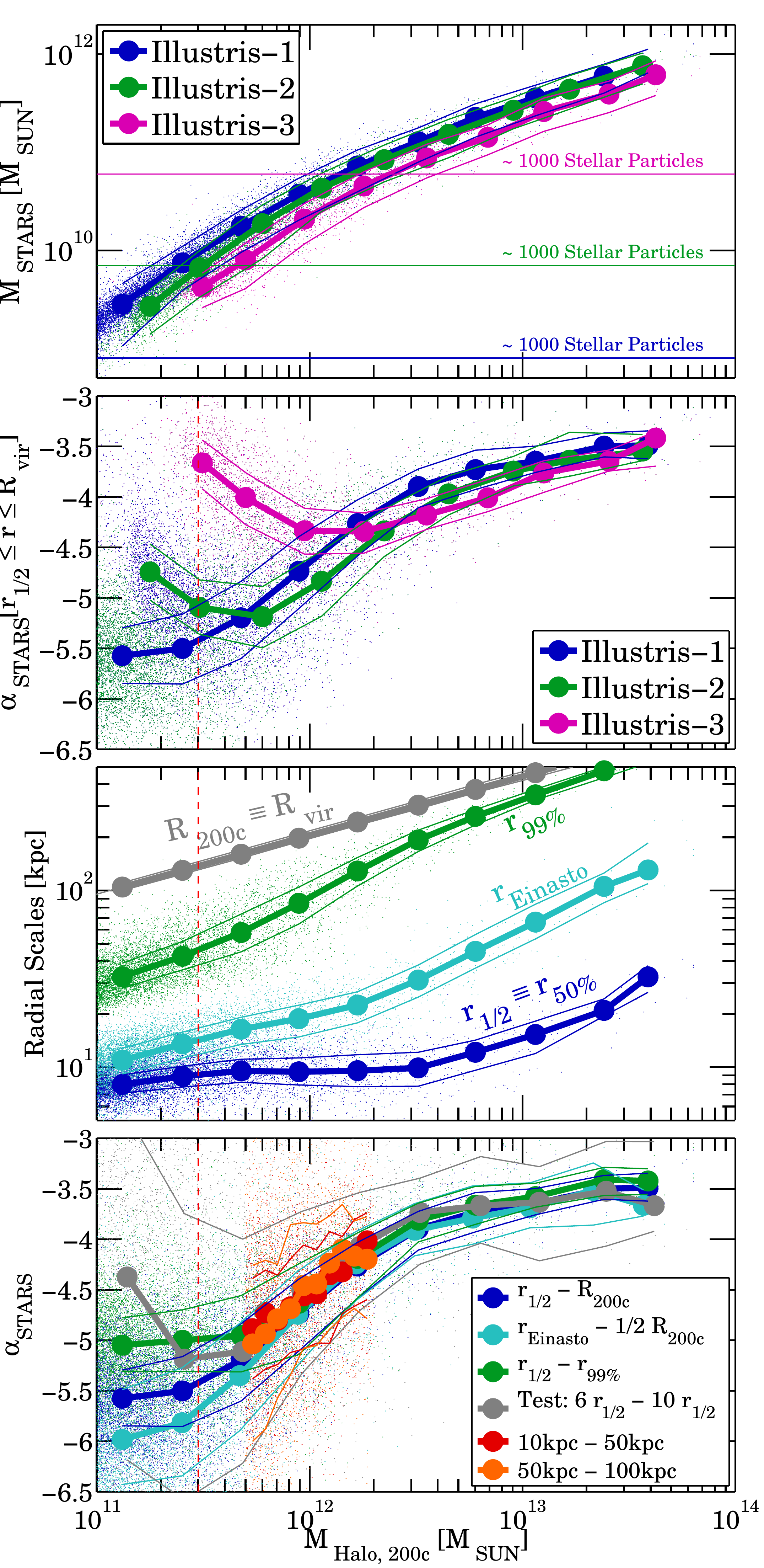}
\caption{Top panel: stellar mass -- halo mass relation of Illustris central galaxies at different levels of resolutions. Second panel from the top: resolution tests in the $\AS-\MHALO$ relation. Lack of resolution (as in Illustris-3) produces an upturn in the slopes of the simulated stellar haloes, however, the result that halo stars fall off much more steeply than the DM densities is robust against numerical artifacts. Third panel from the top: typical sizes of the radial scales adopted in this work as a function of mass: stellar half-mass radius, Einasto scale radius, radius containing 99\% of the stellar mass, and the virial radius, here defined as $R_{200c}$. Bottom panel: dependence of the average $\AS$ on the radial range adopted for the power-law fit. In all panels, small dots represent individual haloes measurements; thin solid curves represent the 25 and 75 percentiles around the median values at fixed halo mass (large, filled circles). The halo-to-halo variation at fixed halo mass in the $\AS$ measurement increases for progressively smaller extents of the considered radial ranges. For Milky-Way mass haloes, among other choices, we have also measured the stellar halo slope at fixed apertures, $10 \le r \le 50$ kpc and $50 \le r \le 100$ kpc, for comparison with observations.}
\label{FIG_APP}
\end{center}
\end{figure}

We give more details here about the mass and radial scales adopted throughout the work, and the results of our convergence tests. 
In Figure \ref{FIG_APP}, from top to bottom, we show: 1) the stellar mass--halo mass relation recovered in Illustris, for different resolution levels; 2) the results of our convergence tests within the Illustris Simulation Suite in terms of the relation between $\AS$ and $\MHALO$; 3) the magnitude of the stellar half-mass radius, the Einasto scale radius, the radius enclosing 99\% of the total stellar mass, and the virial radius as a function of halo mass; 4) the $\AS-\MHALO$ for different choices of the radial extent across the stellar halo adopted for the measurement of the slope.

As it can be seen in the second panel of Figure \ref{FIG_APP}, lack of resolution underestimates $|\AS|$ at the low-mass end. The upturn at the low mass end appears to occur chiefly because the low-density stellar outskirts are not well sampled when an insufficient number of stellar particles are present. We have proved this by testing our measurements when only a quarter of randomly-selected stars are used to sample the stellar halo profiles in Illustris-1, across the usual radial extent $r_{1/2} \le r \le \RVIR$: the median trend results to be perfectly coincident with the blue solid curve down to $3-5 \times 10^{11} \MSUN$, below which it exhibits an upturn similar to the one of the lower resolution levels. Our main results appear to be robust against numerical limitations at all masses above $\sim 3\times 10^{11} \MSUN$, which we thereby adopt as our resolution limit. Above Milky-Way masses, the poorer resolution runs produce slightly steeper stellar halo profiles than Illustris-1: this effect appears less pronounced in the two zoom-in runs of \cite{Marinacci:2014} which have been run at different resolutions (see green data points in Figure \ref{FIG_STELLARSLOPES}); however it might justify why the zoom-in data points tend to be slightly higher than Illustris-1's in the $\AS-\MHALO$ diagram of Figure \ref{FIG_STELLARSLOPES}.

As we show in the bottom panel of Figure \ref{FIG_APP}, the median trend of Figure \ref{FIG_STELLARSLOPES} appears also to be only mildly sensitive to the specific choice of the radial range over which the stellar halo is fitted with a single power-law formula. Indeed, the definition itself of stellar halo is ambiguous, and no single choice for the inner and outer boundaries for the measurement of $\AS$ is ideal across all available halo masses and for all galaxy morphologies. We have tested that the slopes of the stellar haloes as a function of halo mass are consistent, within the 1-sigma variations of Figure \ref{FIG_STELLARSLOPES} and down to the resolution limit, for all the following choices of radial ranges:
\begin{itemize}
\item $\RHALF \leq r \leq \RVIR$,
\item $2\times \RHALF \leq r \leq \RVIR$,
\item $2\times \RHALF \leq r \leq 10\times \RHALF$,
\item $6\times \RHALF \leq r \leq 10\times \RHALF$,
\item $r_{\rm Einasto} \leq r \leq \RVIR$,
\item $r_{\rm Einasto} \leq r \leq 1/2~ \RVIR$, 
\item $\RHALF \leq r \leq r_{90\%}$,
\item $\RHALF \leq r \leq r_{99\%}$, and
\item $10 \leq r \leq 50 $ kpc and $50 \leq r \leq100 $ kpc,
\end{itemize}
the latter for Milky-Way mass haloes only.
By inspecting the spread identified by the thin, solid curves of Figure \ref{FIG_APP} (25 and 75 percentiles around the median values at fixed halo mass), the halo-to-halo variation at fixed halo mass in the $\AS$ measurement increases for progressively smaller extents of the considered radial ranges: this is due to the occurrence of local deviations in the stellar density profiles from pure power-laws.


\small{  

}


\end{document}